\documentclass[twocolumn,english,showpacs]{revtex4}
\usepackage{pslatex}
\usepackage[T1]{fontenc}
\usepackage[latin1]{inputenc}
\usepackage{graphicx}
\usepackage{amssymb}

\makeatletter

\usepackage{graphicx}
\usepackage{babel}
\makeatother
\begin{document}

\title{Exponentially decaying correlations in a gas of strongly interacting
spin-polarized 1D fermions with zero-range interactions}

\author{Scott A. Bender}

\affiliation{Department of Physics, Santa Clara University, Santa Clara, CA 95053}

\author{Kevin D. Erker}

\affiliation{Department of Physics, Santa Clara University, Santa Clara, CA 95053}

\author{Brian E. Granger}

\affiliation{Department of Physics, Santa Clara University, Santa Clara, CA 95053}

\begin{abstract}
We consider the single particle correlations and momentum distributions
in a gas of strongly interacting spinless 1D fermions with zero-range
interactions. This system represents a fermionic version of the Tonks-Girardeau
gas of impenetrable bosons as it can be mapped to a system of noninteracting
1D bosons. We use this duality to show that the $T=0$ single particle
correlations exhibit an \emph{exponential decay} with distance. This
strongly interacting system is experimentally accessible using ultracold
atoms and has a Lorentzian momentum distribution at large momentum
whose width is given by the linear density. 
\end{abstract}


\maketitle
One dimensional (1D) many body systems provide a beautiful example
of duality. In this context, duality refers to the precise mappings
that exist between weakly interacting 1D bosons and strongly interacting
1D fermions and vice versa. A number of classic examples of duality
in 1D systems exist: i) the Luttinger liquid \cite{beg-Tomonaga1950a,beg-Luttinger1963a}
which is dual to a collection of noninteracting bosonic collective
modes, ii) the Tonks-Girardeau gas of impenetrable bosons \cite{beg-Tonks1936a,beg-Girardeau1960a},
which is dual to a noninteracting gas of fermions and iii) the quantum
sine-Gordon model, which is dual to the massive Thirring model \cite{beg-Coleman1975a}. 

More recently, a number of groups \cite{beg-Cheon1999a,beg-Albeverio2000a,beg-Granger2003b,beg-Girardeau2003a,beg-Grosse2004a}
have introduced equivalent models of spinless 1D fermions with zero-range
interactions. Again, duality appears in these models as they can all
be mapped to a system of $\delta$-function interacting 1D bosons.
These models contain the familiar bosonic Tonks-Girardeau gas \cite{beg-Tonks1936a,beg-Girardeau1960a}
as a specific case as well as a fermionic version of the Tonks-Girardeau
(TG) gas \cite{beg-Granger2003b,beg-Girardeau2003a} in which strongly
interacting fermions are related to noninteracting bosons. In this
paper we discuss the correlations among these 1D fermions with strong
interactions. 

Ultracold atom gases provide an excellent arena in which to study
1D many body systems with short-range interactions. Already, ultracold
atoms have been confined in quasi-1D (highly elongated) harmonic traps
\cite{beg-Gorlitz2001a} where phase fluctuations in quasi-1D Bose-Einstein
condensates \cite{beg-Dettmer2001a,beg-Richard2003a} and the TG gas
\cite{beg-Paredes2004a,beg-Kinoshita20004a} have been observed. 

Although such experiments are actually three dimensional (3D) in nature,
their low energy properties can be modeled by an effective 1D Hamiltonian,
as first demonstrated by Olshanii \cite{beg-Olshanii1998a}. For quasi-1D
bosons, this model consists of mass $m$ 1D bosons interacting through
the potential $V(z)=g_{1D}^{B}\delta(z)$, where the interaction strength,
$g_{1D}^{B}$, is chosen to correctly reproduce the low energy behavior
of the full 3D Hamiltonian of the harmonically confined atoms. Most
importantly, $g_{1D}^{B}$ can be tuned to any value (even infinite)
by changing ratio $a_{s}/a_{\perp}$, where $a_{s}$ is the 3D $s$-wave
scattering length of the bosons and $a_{\perp}$ is the length scale
of the tight harmonic confinement. Thus, ultracold atoms in quasi-1D
traps can be used to probe the well studied uniform 1D Bose gas with
$\delta$-function interactions \cite{beg-Lieb1963a,beg-McGuire1964a,geb-Yang1969a}.
The dimensionless interaction strength for this system is $\gamma_{B}=\frac{mg_{1D}^{B}}{n\hbar^{2}}$,
where $n$ is the linear density. In the strongly interacting limit
($\gamma_{B}\rightarrow\infty$) the $\delta$-interacting 1D Bose
gas reduces to the recently observed TG gas \cite{beg-Olshanii1998a,beg-Dettmer2001a,beg-Kinoshita20004a}. 

Of critical importance in these studies are the static correlation
functions. These determine experimentally measurable properties such
as momentum distributions \cite{beg-Paredes2004a} and inelastic collision
rates \cite{beg-Inouye1998a}. For the $\delta$-interacting Bose
gas, these correlation functions are known in a number of different
regimes. In the TG regime ($\gamma_{B}\rightarrow\infty$) duality
can be used to calculate the single particle correlations for all
densities and distances \cite{beg-Vaidya1978a,beg-Olshanii1998a}.
For finite interactions $\gamma_{B}$ the single- and local multi-
particle correlation functions are known analytically in a number
of limits (see \cite{beg-Gangardt2003a,beg-Kheruntsyan2003a,beg-Olshanii2003a}
and the references therein). Lastly, numerical calculations have covered
all interaction strengths for both uniform and harmonically confined
systems \cite{beg-Astrakharchik2002b}. The most important feature
in all these studies is that single particle correlations at zero
temperature ($T=0$) exhibit a \emph{power law decay} with distance.
This power law decay of correlations is also shared by the classic
Luttinger liquid model of spinless 1D fermions \cite{beg-Haldane1981a}. 

With this background we turn our attention to the single particle
correlations in the recently proposed model of spinless 1D fermions
with zero-range interactions. Our focus is upon strongly interacting
1D fermions, which are dual to noninteracting bosons; a sort of fermionic
version of the TG gas. We emphasize that this system does not appear
to be another obvious manifestation of the Luttinger liquid. This
becomes evident in the nature of the single particle correlations,
which we find decay \emph{exponentially} with distance at $T=0$. 

Our method consists of using duality to obtain a general relationship
between the single particle density matrices of the fermions and their
dual bosons. Thus, once the single particle correlations are known
for noninteracting bosons in any external potential, the fermionic
correlations are known immediately. We apply these general results
to uniform and harmonically confined 1D gases, and give the experimentally
relevant momentum distributions for each. 

\emph{Spin polarized fermions.} Here we introduce and summarize the
main features of the 1D spinless fermions considered in this paper.
This model has been introduced by a number of researchers in different
forms \cite{beg-Cheon1999a,beg-Albeverio2000a,beg-Granger2003b,beg-Girardeau2003a,beg-Grosse2004a}.
Consider a Hamiltonian for $N$ mass $m$ 1D spinless fermions,\begin{equation}
H_{F}=\frac{-\hbar^{2}}{2m}\sum_{i=1}^{N}\frac{\partial^{2}}{\partial z_{i}^{2}}+\sum_{i<j}^{N}V_{F}\left(\left|z_{i}-z_{j}\right|\right),\label{eq:1dbosonham}\end{equation}
with a zero-range interaction potential $V_{F}(z)$ that has a nontrivial
action on the space of antisymmetric wavefunctions $\psi_{F}(x_{1},\ldots,x_{N})$.
The zero-range interaction potential $V_{F}(z)$, parameterized by
an interaction strength $g_{1D}^{F}$, can be chosen in a number of
ways, including \cite{beg-Cheon1999a}

\begin{equation}
V_{F}(z)=\lim_{s\rightarrow0}\frac{-\hbar^{2}}{2\mu}\left(\frac{\hbar^{2}}{\mu g_{1D}^{F}}+\frac{1}{s}\right)\left[\delta(z-s)+\delta(z+s)\right]\label{eq:potential1}\end{equation}
and \cite{beg-Kanjilal2004a}\begin{equation}
V_{F}(z)=g_{1D}^{F}\frac{\overleftarrow{\partial}}{\partial z}\delta(z)\frac{\overrightarrow{\partial}}{\partial z}.\label{eq:potential2}\end{equation}
At first glance, it is not obvious that these and the other \cite{beg-Albeverio2000a,beg-Grosse2004a,beg-Girardeau2003a,beg-Olshanii2004a}
zero-range potentials for 1D spinless fermions are equivalent. However,
when treated \emph{exactly}, they all lead to the same logarithmic
derivative of the relative wavefunction $\psi_{rel}(z)$ where two
particles touch ($z$ is the distance between the particles) \cite{beg-Cheon1999a}:\begin{equation}
\left[\frac{d\psi_{rel}(z)}{dz}\frac{1}{\psi_{rel}(z)}\right]_{z\rightarrow0^{+}}=-\frac{\hbar^{2}}{\mu g_{1D}^{F}}.\label{eq:logderiv}\end{equation}
The dimensionless interaction strength for this model is $\gamma_{F}=\frac{mg_{1D}^{F}n}{\hbar^{2}}$
so that weak interactions ($\gamma_{F}\ll1$) occur for low density
and small $g_{1D}^{F}$ and strong interactions ($\gamma_{F}\gg1$)
occur at high density and large $g_{1D}^{F}$. 

Recently, one of us (B. Granger), in collaboration with D. Blume,
showed that this model of 1D fermions with zero-range interactions
describes the physics of spin-polarized fermionic atoms in quasi-1D
harmonic traps at ultracold temperatures \cite{beg-Granger2003b}.
Then, the effective 1D interaction strength $g_{1D}^{F}$ depends
on the ratio $V_{p}/a_{\perp}^{3}$, where $V_{p}$ is the $p$-wave
scattering volume of the atoms and $a_{\perp}$ the length scale of
the tight harmonic confinement. Thus, by changing either $V_{p}$
(using a Feshbach resonance \cite{beg-Regal2003a}) or $a_{\perp}$
(by modifying the trap) $g_{1D}^{F}$ can be tuned over a large range
of values, including infinity. Thus, it should be possible to study
strongly interacting 1D fermions using ultracold atoms; just as strongly
interacting 1D bosons have been studied in recent experiments \cite{beg-Paredes2004a,beg-Kinoshita20004a}.

To calculate the single particle correlations in the strongly interacting
limit ($\gamma_{F}\rightarrow\infty$) we map the 1D fermions to a
system of 1D bosons with reciprocal interaction strength. This mapping
was first used by Girardeau \cite{beg-Girardeau1960a} to relate impenetrable
bosons to noninteracting fermions. Recently, the mapping has been
extended to relate 1D fermions with \emph{any interaction strength}
to 1D bosons with $\delta$-function interactions \cite{beg-Cheon1999a,beg-Albeverio2000a,beg-Granger2003b,beg-Girardeau2003a,beg-Grosse2004a}.
More specifically, \emph{any} antisymmetric wavefunction $\psi_{F}(z_{1},\ldots,z_{N})$
for $N$ 1D fermions with zero-range interactions of strength $g_{1D}^{F}$
is related to a symmetric wavefunction $\psi_{B}(z_{1},\ldots,z_{N})$
of $N$ 1D bosons with $\delta$-function interactions of strength
$g_{1D}^{B}$ through the mapping \cite{beg-Girardeau1960a}: \begin{equation}
\psi_{F}(z_{1},\ldots,z_{N})=\left[\prod_{j<k}^{k=N}sgn(z_{j}-z_{k})\right]\psi_{B}(z_{1},\ldots,z_{N}).\label{eq:fermionwave}\end{equation}
When the interaction strengths of the bosons and fermions are chosen
to be reciprocals of each other,

\begin{equation}
g_{1D}^{F}=\frac{-\hbar^{4}}{\mu^{2}g_{1D}^{B}},\label{eq:strengthrelat}\end{equation}
the energy spectrum, thermodynamic properties and \emph{local} correlations
(as $\left|\prod_{j<k}^{k=N}sgn(z_{j}-z_{k})\right|^{2}=1$) of the
two systems are identical. Thus, for example, strongly interacting
fermions ($\gamma_{F}\rightarrow\infty$) share these properties with
the noninteracting 1D Bose gas.

Not all properties, however, are shared by the fermions and their
bosonic partners. In fact, any nonlocal correlation function will
be different. Thus, nonlocal correlations provide an important and
unique signature of the actual exchange statistics of the physical
particles. We now derive the single particle correlations and momentum
distributions for the 1D fermions described above in the strongly
interacting limit ($\gamma_{F}\rightarrow\infty$). 

\emph{General form of the density matrix.} For both bosons and fermions,
the single particle reduced density matrix is defined as \begin{eqnarray}
\rho(z,z^{\prime}) & = & N\int_{a}^{b}\psi(z,z_{2},\ldots,z_{N})\psi^{*}(z',z_{2},\ldots,z_{N}) \nonumber \\ 
 & \times & dz_{2}\ldots dz_{N},\label{eq:dmatrixdef}\end{eqnarray}
so that $Tr\left[\rho\right]=N$. The ground state wavefunction, $\psi_{B}$,
for $N$ noninteracting bosons has the familiar form

\begin{equation}
\psi_{B}(z_{1},\ldots,z_{N})=\prod_{i=1}^{N}\phi_{0}(z_{i}),\label{eq:evenwave}\end{equation}
where $\phi_{0}(z_{i})$ is the normalized ground state solution to
the single particle Schrödinger equation in the region $a\leq z\leq b$.
From the form of $\psi_{B}$ in Eq.~(\ref{eq:evenwave}), the reduced
density matrix for noninteracting bosons, $\rho_{B}(z,z')$, is easily
found and quite simple:\begin{equation}
\rho_{B}(z,z')=N\phi_{0}(z)\phi_{0}^{*}(z').\label{eq:bosonmatrix}\end{equation}

Finding the density matrix, $\rho_{F}(z,z')$, for 1D fermions with
strong ($\gamma_{F}\rightarrow\infty)$ zero-range interactions, on
the other hand, is nontrivial, but can be accomplished by using duality.
We construct the interacting fermion wavefunction in the $\gamma_{F}\rightarrow\infty$
limit using the noninteracting boson wavefunction, Eq. (\ref{eq:evenwave}),
and the mapping, Eq. (\ref{eq:fermionwave}). Then, substitution of
the fermion wave function $\psi_{F}$, Eq.~(\ref{eq:fermionwave}),
into the definition of the density matrix, Eq.~(\ref{eq:dmatrixdef}),
yields an integral expression for the single particle fermionic density
matrix in terms of the orbitals of the noninteracting bosons:\begin{eqnarray}
\rho_{F}(z,z^{\prime}) & = & N\phi_{0}(z)\phi_{0}^{*}(z')\int_{a}^{b}\left(\prod_{i=2}^{N}sgn(z-z_{i})sgn(z'-z_{i})\right)\nonumber \\
 & \times & \left(\prod_{1<j<k}^{k=N}\left(sgn(z_{j}-z_{k})\right)^{2}\phi_{0}(z_{k})\phi_{0}^{*}(z_{k})\right) \nonumber \\
 & \times & dz_{2}\ldots dz_{N}.\label{eq:fermionmatrix1}\end{eqnarray}
Noticing that $\left(sgn(z_{j}-z_{k})\right)^{2}=1$ in this expression,
the multidimensional integrals can be separated into a product of
$N-1$ integrals over each $dz_{i}$. A careful treatment of these
integrals gives the final expression for the single particle reduced
density matrix for 1D fermions in the strongly interacting ($\gamma_{F}\rightarrow\infty$)
limit: \begin{equation}
\rho_{F}(z,z')=\rho_{B}(z,z')(1-2P(z,z'))^{N-1}.\label{eq:mapping}\end{equation}

In this expression\begin{equation}
P(z,z')=\frac{1}{N}\left|\int_{z}^{z'}\rho_{B}(z'',z'')dz''\right|\label{eq:prob}\end{equation}
 is the probability of finding a noninteracting boson between $z\textrm{ }$and
$z'$. This result gives a completely general relationship between
the single particle density matrices of noninteracting bosons and
strongly interacting fermions in 1D.

In many cases it is possible to take the thermodynamic limit of these
results ($N\rightarrow\infty$, $L\rightarrow\infty$ but $n=N/L$
fixed, where $L$ is some characteristic length for the system). If
$\tilde{P}(z,z')=LP(z,z')$ is independent of $N$ and $L,$ then
the expression\begin{equation}
e^{x}=\lim_{N\rightarrow\infty}\left(1+x/N\right)^{N}\label{eq:explimit}\end{equation}
can be used to take the thermodynamic limit of the fermion density
matrix, Eq.~(\ref{eq:mapping}):

\begin{equation}
\rho_{F}=\rho_{B}e^{-2n\tilde{P}(z,z^{\prime})}.\label{eq:rhofthermo1}\end{equation}
Typically $\tilde{P}(z,z')\sim\left|z-z'\right|$, so that the 1D
fermions generally exhibit an \emph{exponential decay} of correlations
with distance compared to noninteracting bosons. 

\emph{Periodic system.} Now that we have the general expression, Eqs.~(\ref{eq:mapping})
and (\ref{eq:prob}), for the single particle density matrix in the
limit $\gamma_{F}\rightarrow\infty$, we can apply this result to
specific external potentials. First, consider a constant potential
with periodic boundary conditions over the interval $[0,L]$. The
ground state solution for each orbital is just $\phi_{0}(z_{j})=\sqrt{\frac{1}{L}}$,
so from Eq.~(\ref{eq:bosonmatrix}), $\rho_{B}=n$. The fermion density
matrix then follows from Eqs.~(\ref{eq:mapping}) and (\ref{eq:prob}):\begin{equation}
\rho_{F}(z,z')=n(1-2\mid\frac{z'-z}{L}\mid)^{N-1}.\label{eq:rho1N}\end{equation}

The thermodynamic limit ($N\rightarrow\infty$, $L\rightarrow\infty$
and $n=N/L$ fixed) of this result,\begin{equation}
\rho_{F}(z,z')=ne^{-2n\mid z'-z\mid},\label{eq:rho1tdl}\end{equation}
 exhibits the promised exponential decay with distance. Alternatively,
the single particle correlations can be characterized by the correlation
function, $g_{1}(z)=\rho_{F}(z,0)/n=\exp(-2n\left|z\right|),$ which
is normalized as $g_{1}(0)=1$. This exponentially decaying single
particle correlation function is one of the central results of this
paper and clearly shows the striking effect that the boson$\rightarrow$fermion
mapping, Eq.~(\ref{eq:fermionwave}), has on the nature of single
particle correlations. More specifically, the long-range order of
$T=0$, noninteracting 1D bosons becomes an exponential decay of correlations
(with correlation length $\xi=\frac{1}{2n}$) in the strongly interacting
1D fermions considered here. Furthermore, this behavior is a dramatic
departure from the power law decay of correlations exhibited by the
$T=0$ $\delta$-interacting 1D Bose gas and Luttinger liquid.

Of particular importance for experiments with ultracold atoms is the
momentum distribution $n(k)$, which is obtained as the Fourier transform
of the correlation function $g_{1}(z):$\begin{eqnarray}
n(k) & = & \frac{1}{2\pi}\int_{-\infty}^{\infty}e^{ikz}g_{1F}(z)dz\label{eq:momentum}\\
 & = & \frac{2n}{\pi(4n^{2}+k^{2})}.\label{eq:norm}\end{eqnarray}
This Lorentzian momentum distribution shows the signature effect of
strong interactions in a system of zero-temperature fermions. That
is, interactions have completely smoothed out the sharp decrease of
the momentum distribution at the Fermi momentum $k_{F}\approx2\pi n$.

It is also interesting to note that the width of this momentum distribution
is given by the linear density $n$. As $n\rightarrow0$ this momentum
distribution becomes identical to that of a noninteracting gas of
zero-temperature 1D bosons, $\delta(k)$. This is quite an unexpected
momentum distribution for a strongly interacting gas of fermions.
While it is tempting to speculate about the physical meaning of this
feature (is it a noninteracting Bose gas of molecules?), we warn that
our results are only relevant in the high density limit ($\gamma_{F}\rightarrow\infty$)
limit. 

\emph{Harmonic confinement.} Finally, we can determine the density
matrix and momentum distribution for $N$ 1D fermions of mass $m$
confined to a harmonic oscillator of frequency $\omega_{z}$ in the
$z$ direction. The normalized orbitals are \begin{equation}
\phi_{0}(z_{j})=\sqrt{\frac{1}{a_{z}\sqrt{\pi}}}e^{-z_{j}^{2}/2a_{z}^{2}},\label{eq:oscillatororbital}\end{equation}
 where $a_{z}=\sqrt{\hbar/m\omega_{z}}$. The density matrix for noninteracting
1D bosons, Eq.~(\ref{eq:bosonmatrix}), is then \begin{equation}
\rho_{B}=ne^{\frac{-z^{2}-z'^{2}}{2a_{z}^{2}}},\label{eq:bmtrxoscill}\end{equation}
where we define the linear density for the harmonic oscillator as
$n=\frac{N}{a_{z}\sqrt{\pi}}$. The density matrix for our strongly
interacting 1D fermions, Eqs.~(\ref{eq:mapping}) and (\ref{eq:prob}),
can be written in terms of $Erf(x)=\frac{2}{\sqrt{\pi}}\int_{0}^{x}e^{-t^{2}}dt$:
\begin{equation}
\rho_{F}(z,z')=\rho_{B}(1-\mid Erf(\frac{z}{a_{z}})-Erf(\frac{z'}{a_{z}})\mid)^{N-1}.\label{eq:fermmatrxoscill}\end{equation}
 From this expression, we calculate the momentum distribution $n(k)$
numerically, by Fourier transforming the density matrix,\begin{equation}
\rho_{F}(k,k')=\frac{1}{2\pi}\int_{-\infty}^{\infty}\int_{-\infty}^{\infty}e^{ikz}e^{-ik'z'}\rho(z,z')dzdz',\label{eq:transform}\end{equation}
so that $n(k)=\rho_{F}(k,k)$.

\begin{figure}
\includegraphics[%
  clip,
  width=0.80\columnwidth,
  keepaspectratio,
  angle=270]{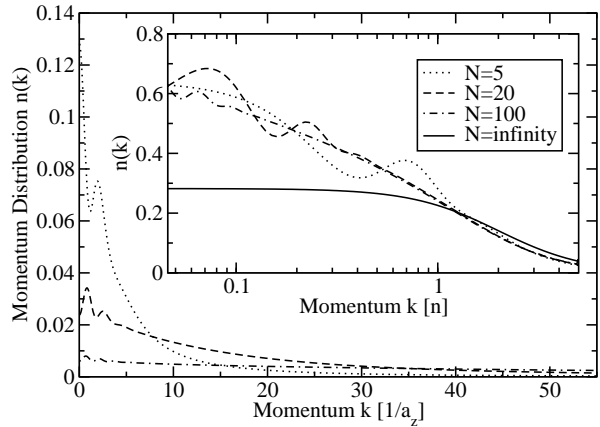}

\caption{\label{fig:fig1}Momentum distributions $n(k)$ are shown as a function
of $k$ for a 1D gas of strongly interacting spinless fermions with
1D harmonic confinement of frequency $\protect\omega_{z}$. The main
figure shows the momentum distributions, normalized to unity with
the momentum given in units of $1/a_{z}$. The inset shows the momentum
distribution (normalized to $N$) with the momentum given in units
of the density $n=N/a_{z}\protect\sqrt{\pi}$. For the $N=\infty$
case we have plotted the Lorentzian, Eq.~(\ref{eq:norm}), that corresponds
to the thermodynamic limit. }
\end{figure}

Figure \ref{fig:fig1} shows the calculated momentum distributions
of harmonically confined, strongly interacting 1D fermions for $N=5$,
$20$ and $100$ particles. The main figure shows clearly that the
widths of these distributions increase in proportion with the linear
density $n$. This is the opposite behavior from what has been calculated
for harmonically confined 1D interacting bosons \cite{beg-Astrakharchik2002b}.
The oscillations at small momentum are due to the harmonic confinement
rather than the details of the interactions. 

At a more quantitative level, the large momentum tails of these distributions
can be understood by taking the thermodynamic limit of the density
matrix, Eq.~(\ref{eq:fermmatrxoscill}). This limit is specified
by letting $a_{z}\sqrt{\pi}\rightarrow\infty$ and $N\rightarrow\infty$
with $n=N/a_{z}\sqrt{\pi}$ fixed. In this limit, we can use the small
argument approximation $Erf(\frac{z}{a})\approx\frac{2z}{a_{z}\sqrt{\pi}}$
and Eq.~(\ref{eq:explimit}) to obtain $\rho_{F}=\rho_{B}\exp(-2n\left|z-z'\right|)$.
Thus, in the thermodynamic limit, the short range correlations in
a harmonic potential are identical to those of a periodic system.
In the momentum distributions, this shows up as a Lorentzian tail,
Eq.~(\ref{eq:norm}), with a width of $4n$, where $n$ is the linear
density. This feature is confirmed in the inset of Fig.~\ref{fig:fig1},
where the momentum is plotted on a logarithmic scale in units of the
density $n$ to show how at large momentum all the distributions line
up with a Lorentzian of width $4n$ (solid line). This Lorentzian
momentum tail will be a clear experimental signature of the strongly
interacting regime discussed here.

In conclusion, duality in 1D many body systems provides a powerful
method for understanding and calculating correlations in the strongly
interacting limit. The system we consider, namely, strongly interacting
spin-polarized 1D fermions with zero-range interactions, is dual to
a noninteracting gas of bosons and provides an excellent prospect
to study duality in actual experiments with ultracold atom gases.
The exponentially decaying correlations described here will have a
clear and measurable effect in these experiments and will hopefully
spur on further theoretical inquiries as well. 

This work was funded by the NSF under grant PHY-0354882. 

\bibliographystyle{apsrev}
%\bibliography{/Users/bgranger/Desktop/1dcorr/1dscat}

\end{document}